\def\cm2{cm$^{-2}$}

\def\c2{C~{\sc ii}}
\def\c4{C~{\sc iv}}
\def\fe2{Fe~{\sc ii}}
\def\fe3{Fe~{\sc iii}}
\def\mg1{Mg~{\sc i}}
\def\mg2{Mg~{\sc ii}}
\def\si2{Si~{\sc ii}}
\def\si4{Si~{\sc iv}}
\def\al2{Al~{\sc ii}}
\def\al3{Al~{\sc iii}}
\def\o1{O~{\sc i}}
\def\n1{N~{\sc i}}
\def\h1{H~{\sc i}}

\def\approxlt{\mathrel{\spose{\lower 3pt\hbox{$\sim$}}
        \raise 2.0pt\hbox{$<$}}}
\def\approxgt{\mathrel{\spose{\lower 3pt\hbox{$\sim$}}
        \raise 2.0pt\hbox{$>$}}}

\documentclass[article]{gwp80}
\usepackage{graphicx}
\usepackage{amssymb}
\tabletypesize{\normalsize}  

\shortauthors{Kanbur et al}
\shorttitle{Galactic Oosterhoff Dichotomy}

\begin{document}
\large    
\pagenumbering{arabic}
\setcounter{page}{112}

\title{The Galactic Oosterhoff Dichotomy in terms of\\ \\ Period-Color Relations at Maximum/Minimum\\ \\ Light}

%
%
\author{{\noindent Sashi Kanbur,{$^{\rm 1}$} Ata Sarajedini,{$^{\rm 2}$},
Karen Kinemuchi {$^{\rm 3}$}, Roger Cohen,{$^{\rm 2}$} and \\ Chow Choong Ngeow{$^{\rm 4}$}\\
\\
{\it (1) SUNY Oswego, NY, USA\\
(2) University of Florida, USA\\
(3) NASA-Ames Research Center/Bay Area Environmental Research Institute\\
(4) National Central University, Taiwan}
}
}

%
%
\email{(1) shashi.kanbur@oswego.edy}


\begin{abstract}
The Galactic Oosterhoff dichotomy between Galactic globular clusters of type I (OoI) and
type II (OoII) is often characterized by a difference in the period-amplitude (PA) relations displayed
by RR Lyrae ab stars in each type of cluster. Classical examples of OoI and OoII are M3 and M15 respectively.
Here we use multicolor data for the these two clusters to demonstrate how period-color (PC) relations at maximum
$V$ band light are also different in OoI and OoII clusters.   
\end{abstract}

\section{Introduction}

Fundamental mode RRab stars in Galactic globular clusters of type I and type II have an average period of
about 0.55 and 0.65 days respectively. This is the standard way of characterizing the Galactic Oosterhoff
dichotomy. Another approach (see Cacciari et al (2005) and references therein) looks at PA relations. RRab stars in OoI clusters like M3 have
very different PA relations when compared to RRab stars in OoII clusters like M15. 

The data in this paper are taken from Benko et al (2006) and from Corwin et al (2008). We used a value of
E(V-I) of 0.0128 and 0.128 for M3 and M15 respectively. Because of the high quality of these data, we fit a relatively
high oder Fourier fit and used the subsequent Fourier expansion to estimate light curve minimum, maximum and mean.  

In Figure 1, we clearly see
two different PA relations for the two clusters. Moreover, there are a group of stars in M3 which follow the PA relation for M15.
One possible explanation for this is that OoII clusters are more evolved than OoI clusters and those stars in M3 which follow the M15 relation
are also more evolved.

\begin{figure*}
\centering
\includegraphics[width=9cm,angle=270]{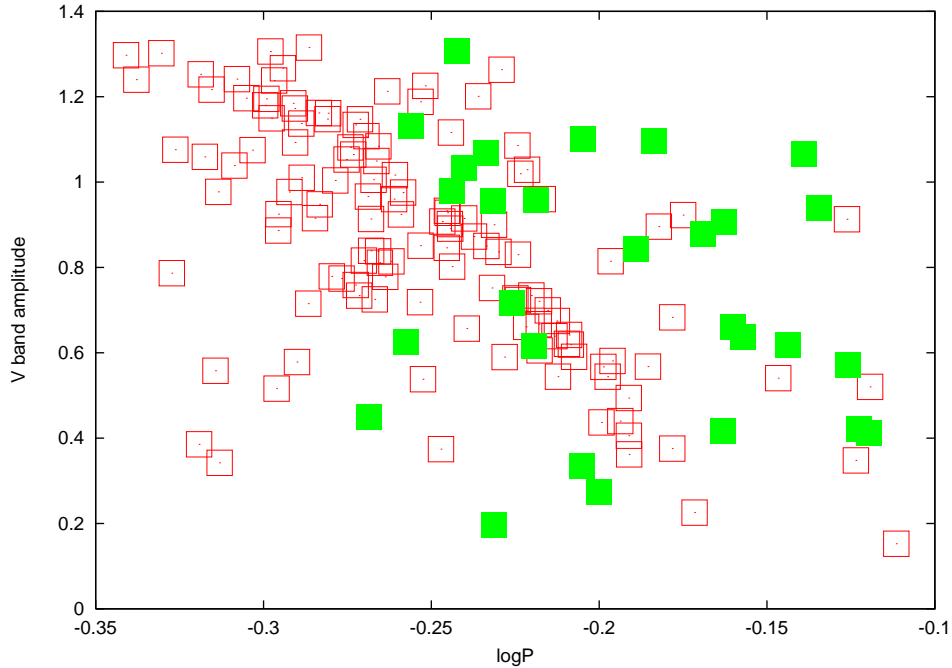}
\vskip0pt
\caption{Log Period against V band amplitude for M3 (open squares) and
M15 (solid squares). Notice the two distinct PA relations for the two clusters and the
fact that some stars in M3 follow the M15 relation.} 
\label{amplogP}
\end{figure*}

\section{Period-Color Relations at Maximum Light}

We use our Fourier fits to construct $(V-I)$ colors at maximum, minimum and mean $V$ band light. That is
we rephase our data so that maximum $V$ band light occurs at phase 0. Then the $I$ band curve is rephased accordingly.
Figure 2 represents a plot of log Period against $V-I$ color at maximum $V$ band light, corrected for
extinction as specified above.

We clearly see two well separated sequences for M3 and M15. The same M3 stars which lie on the M15 PA relation in Figure 1
also lie on the M15 PC maximum light relation in Figure 2.

\begin{figure*}
\centering
\includegraphics[width=9cm,angle=270]{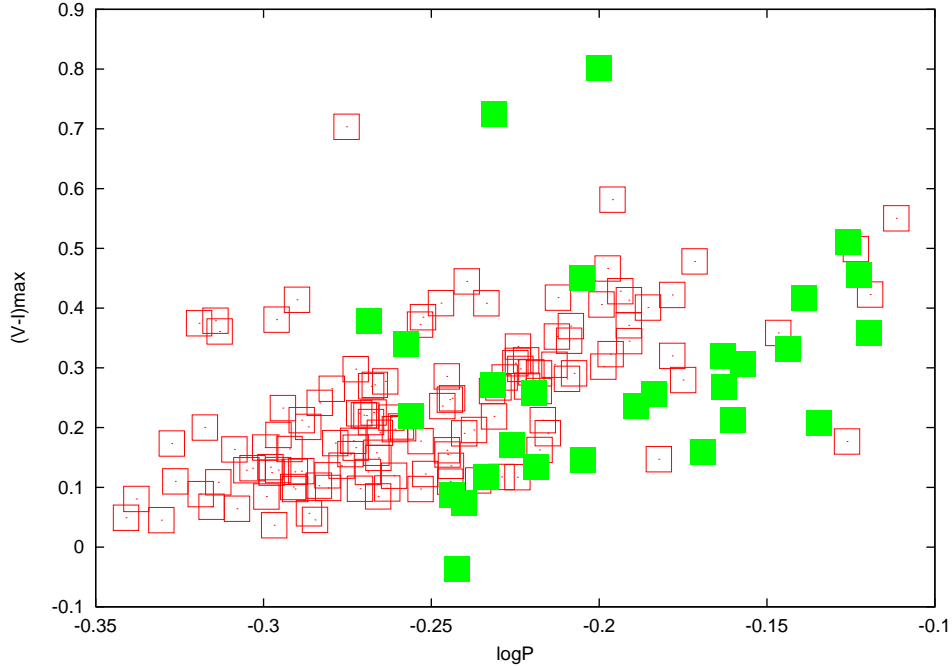}
\vskip0pt
\caption{Log Period against $V-I$ color at maximum $V$ band light for M3 (open squares) and
M15 (solid squares), corrected for extinction. We clearly see similar features to the PA diagram: two different
relations for M3 and M15 and those stars in M3 which follow the M15 PA relation also follow
the M15 PC relation at maximum light.}
\label{vimaxlogP}
\end{figure*}

\vfill\eject

\section{Period-Color Relations at Mean and Minimum Light}

\begin{figure*}[h]
\centering
\includegraphics[width=9cm,angle=270]{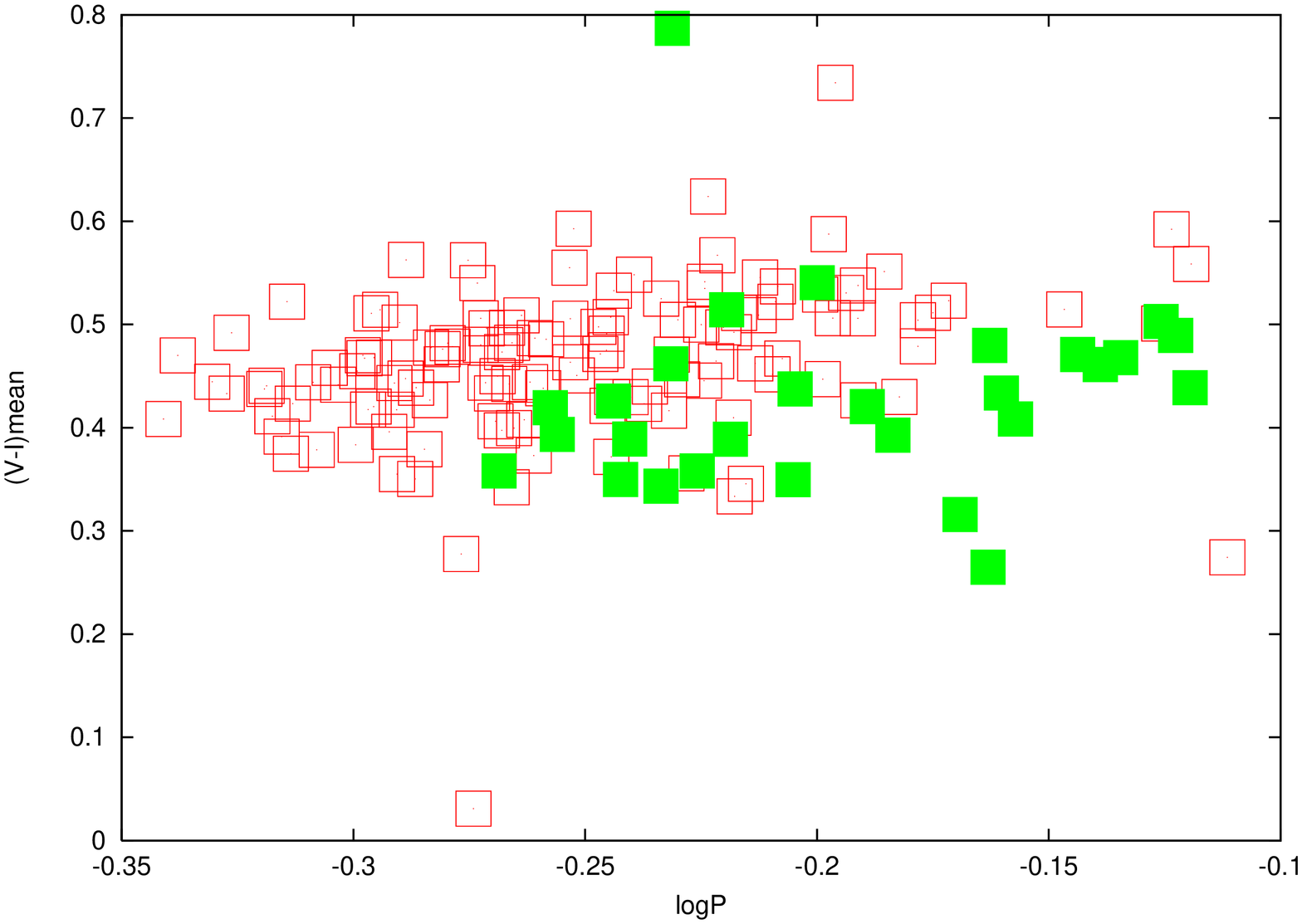}
\vskip0pt
\caption{Log Period against $V-I$ color at mean $V$ band light for M3 (open squares) and
M15 (solid squares), corrected for extinction. Some structure is present but the separation seen in the PA or
PC relation at maximum light is clearly not seen.}
\label{vimeanlogP}
%
%
\centering
\includegraphics[width=9cm,angle=270]{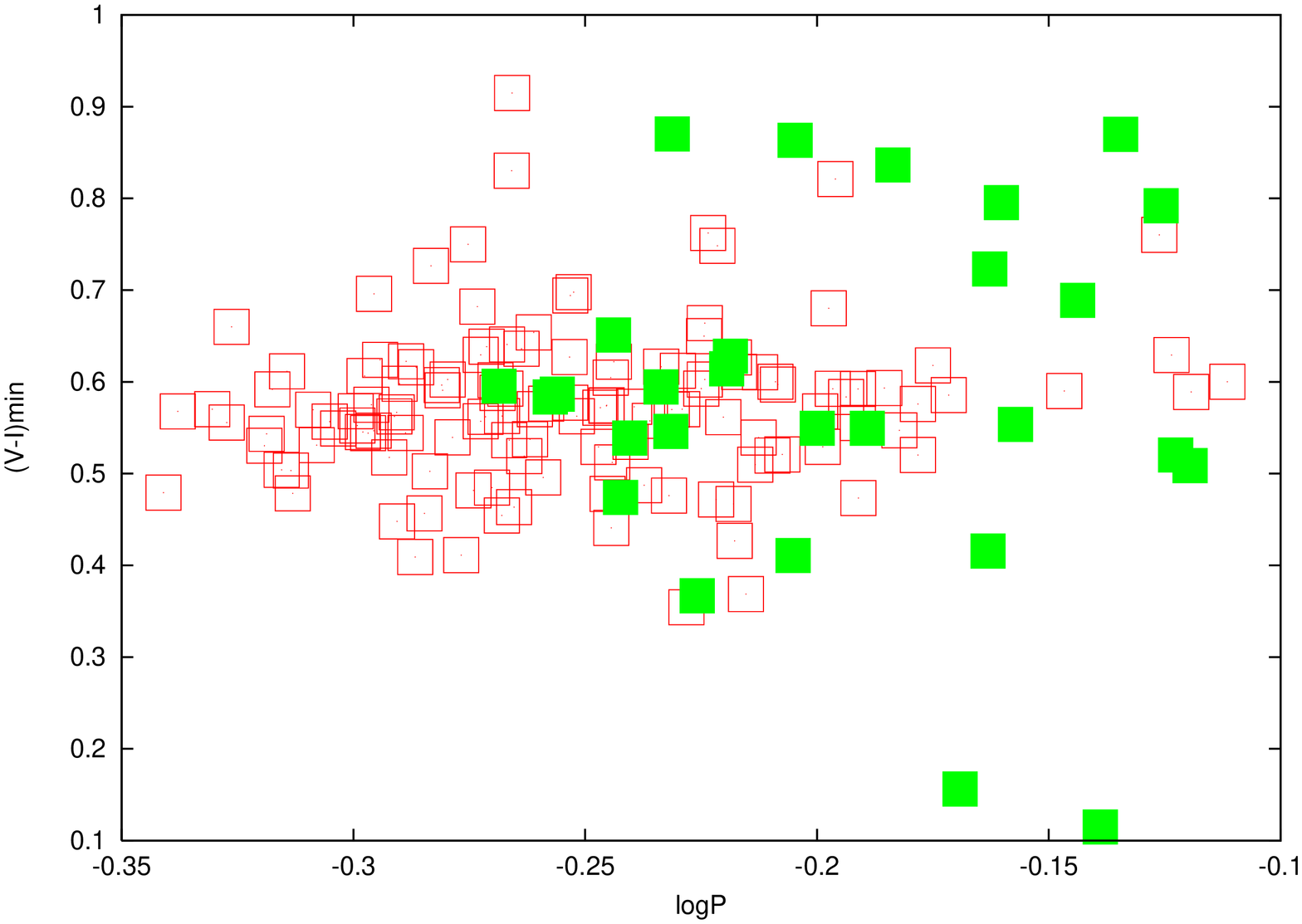}
\vskip0pt
\caption{Log Period against $V-I$ color at minimum $V$ band light for M3 (open squares) and M15
(solid squares), corrected for extinction. We clearly see a flat relation at minimum light for both clusters.}
\label{viminlogP}
\end{figure*}

Figures 3 and 4 display PC relations at mean and minimum light respectively. Figure 3 clearly has lost the 
separation between the two types of clusters demonstrated in Figures 1 and 2. Figure 4 shows a flat PC relation at minimum
light for both types of clusters.

\section{Discussion}

Figures 1 and 2 suggest that the Galactic Oosterhoff dichotomy can be recast in terms of PC relations at maximum light.
Further, the PC relations at maximum light also show evidence of a quadratic nature as has been suggested for the PA relations
(Cacciari et al 2005). One reason why casting the Oosterhoff dichotomy in terms of PC relations at maximum light is that there is some
theoretical work which can provide a possible framework within which to understand the differences in PC relations at maximum light in terms of
an evolutionary context (Kanbur and Phillips 1996).

This separation starts to disappear at mean light and reverts to the flat PC relation at minimum light (Kanbur 1995). These
data suggest that there is no difference in the flat PC relation at minimum light between OoI and OoII clusters.

One caveat is that data from more Galactic globular clusters are needed to further test these hypotheses. The analysis of such observations is
currently under way.

\end{document}